\documentclass[a4paper,envcountsame,envcountsect]{llncs}

\usepackage{emlines}

\begin{document}

\title{Probabilistic Reversible Automata and Quantum Automata}

\author{
Marats Golovkins
\thanks{Research partially supported by the Latvian Council of Science, grant No. 01.0354 and grant for Ph.D. students;
University of Latvia, K. Morbergs grant; European Commission, contract IST-1999-11234} 
and Maksim Kravtsev
\thanks{Research partially supported by the Latvian Council of Science, grant No. 01.0354 and European Commission,
contract IST-1999-11234}
}

\institute{
 Institute of Mathematics and Computer Science,
 University of Latvia\\ Rai\c na bulv. 29, Riga, Latvia\\
\email{marats@latnet.lv, maksims@batsoft.lv}
}

\maketitle

%\special{papersize=8.5in,11in}

\begin{abstract}
To study relationship between quantum finite automata and probabilistic finite automata, we introduce a
notion of probabilistic reversible automata (PRA, or doubly stochastic automata).
We find that there is a strong relationship between different possible models of PRA
and corresponding models of quantum finite automata. We also propose a classification of reversible
finite 1-way automata.
\end{abstract}

\section{Introduction}

Here we introduce common notions used throughout the paper as well as summarize its contents.

We analyze two models of probabilistic reversible automata in this paper, namely,
1-way PRA and 1.5-way PRA.

In this section, we define notions applicable to both models in a quasi-formal way, including
a general definition for probabilistic reversibility. These notions are defined formally in further sections.

If not specified otherwise, we denote by $\Sigma$ an input alphabet of an automaton.

Every input word is enclosed into {\em end-marker} symbols $\#$ and $\$$.
Therefore we introduce a {\em working alphabet} as $\Gamma=\Sigma\cup\{\#,\$\}$.

By $Q$ we normally understand the set of states of an automaton.

By $\overline L$ we understand complement of a language $L$.

Given an input word $\omega$, by $|\omega|$ we understand the number of symbols in $\omega$ and
with $[\omega]_i$ we denote i-th symbol of $\omega$, counting from the beginning (excluding end-markers).

\begin{definition}
A {\em configuration} of a finite automaton is
$c=\left<\nu_i q_j\nu_k\right>$, where the automaton is in a
state $q_j\in Q$, $\nu_i\nu_k\in\#\Sigma^*\$$ is a finite word on the input
tape and input tape head is above the first symbol of the word $\nu_k$.
\end{definition}

By $C$ we denote the set of all configurations of an automaton. This set is countably infinite.

After its every step, a probabilistic automaton is in some probability
distribution $p_0 c_0+p_1 c_1+\ldots+p_n c_n$,
where $p_0+p_1+\ldots+p_n=1$. 
Such probability distribution is called a {\em superposition of configurations}.
Given an input word $\omega$, the number of configurations in every accessible superposition
does not exceed $|Q|$ in case of 1-way automata, and $|\omega||Q|$
in case of 1.5-way automata.

A linear closure of $C$ forms a linear space, where every configuration can be viewed as a basis vector.
This basis is called a {\em canonical basis}.
Every probabilistic automaton defines a linear operator over this linear space.

Let us consider A. Nayak's model of quantum automata with mixed states. (Evolution is 
characterized by a unitary matrix and subsequent measurements are performed after each step, POVM
measurements not being allowed, \cite{N 99}.)
If a result of every measurement is a single configuration, not a superposition, and
measurements are performed after each step, we actually get a probabilistic automaton.
However, the following property applies to such probabilistic automata - their evolution
matrices are {\em doubly} stochastic.
This encourages us to give the following definition for probabilistic reversible automata:
%We propose the following definition for probabilistic reversible automata:
\begin{definition}
\label{PRA-def}
A probabilistic automaton is called {\em reversible} if its linear operator can be described by a doubly stochastic
matrix, using canonical basis.
\end{definition}
To make accessible configurations of type $\left<q_i\#\omega\$\right>$, we assume that every word is written on
a circular tape, and after the right end-marker $\$$ the next symbol is the left end-marker $\#$. Such precondition
is the same as used for quantum finite automata. (See, for example, \cite{KW 97}.)

At least two definitions exist, how to interpret word acceptance, and hence, language recognition, for reversible automata.
\begin{definition}
\label{class-acc-def}
{\em Classical acceptance.}
We say that an automaton accepts (rejects) a word classically,
if its set of states consists of two disjoint subsets: accepting states and rejecting states, and
the following conditions hold:
\begin{itemize}
\item the automaton accepts the word, if it is in accepting state after having read the last symbol of the word;
\item the automaton rejects the word, if it is in rejecting state after having read the last symbol of the word.
\end{itemize}
\end{definition}
We refer to the classical acceptance automata as C-automata further in the paper.
\begin{definition}
\label{dh-acc-def}
{\em ``Decide and halt" acceptance.}
We say that an automaton accepts (rejects) a word in a decide-and-halt manner,
if its set of states consists of three disjoint subsets: accepting states, rejecting states and non-halting states, and
the following conditions hold:
\begin{itemize}
\item the computation is continued only if the automaton enters a non-halting state.
\item if the automaton enters an accepting state, the word is accepted;
\item if the automaton enters a rejecting state, the word is rejected.
\end{itemize}
\end{definition}
We refer to the decide-and-halt automata as DH-automata further in the paper.

Having defined word acceptance, we define language recognition in an equivalent way as in \cite{R 63}.
We consider only bounded error language recognition in this paper.

By $P_{x,A}$ we denote the probability that a word $x$ is accepted by an automaton $A$.

\begin{definition}
\label{rec-def}
We say that a language $L$ is recognized with bounded error by an automaton $A$
with interval $(p_1,p_2)$ if $p_1<p_2$ and $p_1=\sup\{P_{x,A}\ |\ x\notin L\}$, $p_2=\inf\{P_{x,A}\ |\ x\in L\}$.
\end{definition}

\begin{definition}
\label{rec1-def}
We say that a language is recognized with a probability $p$ if
the language is recognized with interval $(1-p,p)$.
\end{definition}

\begin{definition}
\label{rec1eps-def}
We say that a language is recognized with probability $1-\varepsilon$,
if for every $\varepsilon>0$ there exists an automaton which recognizes
the language with interval $(\varepsilon_1,1-\varepsilon_2)$, where $\varepsilon_1,\varepsilon_2\leq\varepsilon$.
\end{definition}

\begin{definition}
By $q\stackrel{S}{\longrightarrow}q^\prime$, $S\subset\Sigma^*$, we denote that there
is a positive probability to get to a state $q^\prime$ by reading a single word $\xi\in S$, starting in a state $q$.

\end{definition}

We refer to several existing models of quantum finite automata:
\begin{enumerate}
\item Measure-once quantum finite automata \cite{MC 00} (QFA-MC);
\item Measure-many quantum finite automata \cite{KW 97} (QFA-KW);
\item Enhanced quantum finite automata \cite{N 99} (QFA-N).
\end{enumerate}

Following the notions above, QFA-MC can be characterized as C-automata whereas QFA-KW and QFA-N as
DH-automata.

In Section \ref{Sec1wayC}, we discuss properties of PRA C-automata (PRA-C).
We prove that PRA-C recognize the class of languages $a_1^*a_2^*\ldots a_n^*$ with probability $1-\varepsilon$.
This class can be recognized by QFA-KW, with worse acceptance probabilities,
however \cite{ABFK 99}. This also implies that QFA-N recognize this class of languages
with probability $1-\varepsilon$.

Further, we show general class of regular languages, not recognizable by \mbox{PRA-C}.
In particular, such languages as (a,b)*a and a(a,b)* are in this class. This class has strong similarities
with the class of languages, not recognizable by QFA-KW \cite{AKV 01}.

We also show that the class of languages recognized by PRA-C is closed under boolean operations, inverse homomorphisms
and word quotient, but is not closed under homomorphisms.

In Section \ref{Sec1wayDH} we prove, that PRA DH-automata do not recognize the language (a,b)*a.

In Section \ref{Sec15way} we discuss some properties of 1.5-way PRA. We also present an
alternative notion of probabilistic reversibility, not connected with quantum automata.

In Section \ref{hierarhija} we propose a classification of reversible automata (deterministic, probabilistic and quantum).

\section{1-way Probabilistic Reversible C-Automata}
\label{Sec1wayC}

\begin{definition}
\label{prac-def}
1-way probabilistic reversible C-automaton (PRA-C)\\ $A=(Q,\Sigma,q_0,Q_F,\delta)$ is specified by a finite 
set of states $Q$, a finite input alphabet $\Sigma$, an initial state $q_0\in Q$,
a set of accepting states $Q_F\subseteq Q$, and a transition function
$$
\delta:Q\times\Gamma\times Q\longrightarrow\bbbr_{[0,1]},
$$
where $\Gamma=\Sigma\cup\{\#,\$\}$ is the input tape alphabet of $A$ and $\#$, $\$$
are end-markers not in $\Sigma$. Furthermore, transition function satisfies the
following requirements:
\begin{eqnarray}
&&\forall(q_1,\sigma_1)\in Q\times\Gamma\ \sum\limits_{q\in Q}\delta(q_1,\sigma_1,q)=1
\label{sc1-1}\\
&&\forall(q_1,\sigma_1)\in Q\times\Gamma\ \sum\limits_{q\in Q}\delta(q,\sigma_1,q_1)=1
\label{sc1-2}
\end{eqnarray}
\end{definition}

For every input symbol $\sigma\in\Gamma$, the transition function may be determined
by a $|Q|\times|Q|$ matrix $V_\sigma$, where $(V_\sigma)_{i,j}=\delta(q_j,\sigma,q_i)$.

\begin{lemma}
All matrices $V_\sigma$ are doubly stochastic iff conditions (\ref{sc1-1})
and (\ref{sc1-2}) of Definition \ref{prac-def} hold.
\end{lemma}
\begin{proof}
Trivial. \qed
\end{proof}

We define word acceptance as specified in Definition \ref{class-acc-def}.
The set of rejecting states is $Q\setminus Q_F$.
We define language recognition as in Definition \ref{rec-def}.

A linear operator $U_A$ corresponds to the automaton $A$.
Formal definition of this operator follows:

\begin{definition}
Given a configuration $c=\langle\nu_i q_j\sigma\nu_k\rangle$,
$$
U_Ac\stackrel{\scriptscriptstyle {\rm def}}{=}\sum\limits_{q\in Q}\delta(q_j,\sigma,q)\langle\nu_i\sigma q\nu_k\rangle.
$$
Given a superposition of configurations $\psi=\sum\limits_{c\in C}p_c c$,
$$
U_A\psi\stackrel{\scriptscriptstyle {\rm def}}{=}\sum\limits_{c\in C}p_c U_A c.
$$
\end{definition}

Using canonical basis, $U_A$ is described by an infinite matrix $M_A$.

To comply with Definition \ref{PRA-def}, we have to state the following:
\begin{lemma}
Matrix $M_A$ is doubly stochastic iff conditions (\ref{sc1-1})
and (\ref{sc1-2}) of Definition \ref{prac-def} hold.
\end{lemma}
\begin{proof}
Condition (\ref{sc1-1}) takes place if and only if the sum of elements in every column in $M_A$ equal to $1$.
Condition (\ref{sc1-2}) takes place if and only if the sum of elements in every row in $M_A$ equal to $1$.
\qed
\end{proof}

This completes our formal definition of PRA-C.

Use of end-markers does not affect computational power of PRA-C. For every PRA-C
with end-markers which recognizes some language it is possible to construct a PRA-C without 
end-markers which recognizes the same language. (Number of states needed may increase, however.)
See Appendix for further details.

\begin{lemma}
\label{cpm-lem}
If a language is recognized by a PRA-C $A$ with interval $(p_1,p_2)$, exists
a PRA-C which recognizes the language with probability $p$, where
$$p=
\left\{
\begin{array}{l}
\frac{p_2}{p_1+p_2},\ \ \ \mbox{if }p_1+p_2\geq1\\
\frac{1-p_1}{2-p_1-p_2},\mbox{if }p_1+p_2<1.
\end{array}
\right.$$
\end{lemma}
\begin{proof}
Let us assume, that the automaton $A$ has $n-1$ states.
We consider the case $p_1+p_2>1$.

Informally, having read end-marker symbol $\#$, we simulate the automaton
$A$ with probability $\frac{1}{p_1+p_2}$ and reject input with probability
$\frac{p_1+p_2-1}{p_1+p_2}$.

Formally, to recognize the language with probability $\frac{p_2}{p_1+p_2}$, we modify
the automaton $A$. We add a new state $q_r\notin Q_F$, and change the transition 
function in the following way:
\begin{itemize}
\item $\forall \sigma$, $\sigma\neq\#$, $\delta(q_r,\sigma,q_r)
  \stackrel{\scriptscriptstyle {\rm def}}{=}1$;
\item $\delta(q_0,\#,q_r)
  \stackrel{\scriptscriptstyle {\rm def}}{=}\frac{p_1+p_2-1}{p_1+p_2}$;
\item $\forall q$, $q\neq q_r$, $\delta(q_0,\#,q)
  \stackrel{\scriptscriptstyle {\rm def}}{=}\frac{1}{p_1+p_2}\delta_{old}(q_0,\#,q)$.
\end{itemize}
Now the automaton has $n$ states.
Since end-marker symbol $\#$ is read only once at the beginning of
an input word, we can disregard the rest of transition function values,
associated with $\#$:
$\forall q_i,q_j$, where $q_i\neq q_0$,
$\delta(q_i,\#,q_j)
\stackrel{\scriptscriptstyle {\rm def}}{=}\frac{1-\delta(q_0,\#,q_j)}{n-1}$.

The transition function satisfies the requirements of Definition \ref{prac-def} and
the constructed automaton recognizes the language with probability
$\frac{p_2}{p_1+p_2}$.

The case $p_1+p_2<1$ is very similar. Informally, having read end-marker 
symbol $\#$, we simulate the automaton $A$ with probability
$\frac{1}{2-p_1-p_2}$ and accept input with probability $\frac{1-p_1-p_2}{2-p_1-p_2}$.
\qed
\end{proof}

\begin{theorem}
\label{1eps-lem}
If a language is recognized by a PRA-C, it is recognized by \mbox{PRA-C} with
probability $1-\varepsilon$.
\end{theorem}
\begin{proof}
We assume that a language $L$ is recognized by a 
PRA-C automaton $A=(Q,\Sigma,q_0,Q_F,\delta)$ with interval $(p_1,p_2)$.
Let $\delta=\frac{1}{2}(p_1+p_2)$.

Let us consider a system of $m$ copies of the automaton $A$, denoted as $A_m$.
We say that our system has accepted (rejected) a word if more (less or equal)
than $m\delta$ automata in the system have accepted (rejected) the word.
We define language recognition as in Definition \ref{rec-def}.

Let us consider a word $\omega\in L$. The automaton $A$ accepts $\omega$ with
probability $p_\omega\geq p_2$. As a result of reading $\omega$,
$\mu_m^\omega$ automata of the system accept the word, and the rest reject it. The
system has accepted the word, if $\frac{\mu_m^\omega}{m}>\delta$. Let us take
$\eta_0$, such that $0<\eta_0<p_2-\delta\leq p_w-\delta$. Estimating the 
probability that $\frac{\mu_m^\omega}{m}>\delta$, we have
\begin{equation}
\label{1eps-lem-f1}
P\left\{\frac{\mu_m^\omega}{m}>\delta\right\}\geq P\left\{p_\omega-\eta_0
<\frac{\mu_m^\omega}{m}<p_\omega+\eta_0\right\}
=P\left\{\left|\frac{\mu_m^\omega}{m}-p_\omega\right|<\eta_0\right\}
\end{equation}
In case of $m$ Bernoulli trials, Chebyshev's inequality may be used to prove the
following (\cite{GS 97}, p. 312):
\begin{equation}
\label{1eps-lem-f2}
P\left\{\left|\frac{\mu_m^\omega}{m}-p_\omega\right|\geq\eta_0\right\}
\leq\frac{p_\omega(1-p_\omega)}{m\eta_0^2}\leq\frac{1}{4m\eta_0^2}
\end{equation}
%(\ref{1eps-lem-f2})
The last inequality induces that
\begin{equation}
\label{1eps-lem-f3}
P\left\{\left|\frac{\mu_m^\omega}{m}-p_\omega\right|<\eta_0\right\}\geq1-\frac{1}{4m\eta_0^2}
\end{equation}
Finally, putting (\ref{1eps-lem-f1}) and (\ref{1eps-lem-f3}) together, 
\begin{equation}
\label{1eps-lem-f4}
P\left\{\frac{\mu_m^\omega}{m}>\delta\right\}\geq1-\frac{1}{4m\eta_0^2}
\end{equation}
Inequality (\ref{1eps-lem-f4}) is true for every $\omega\in L$.

On the other hand, let us consider a word $\xi\notin L$. The automaton $A$ accepts $\xi$
with probability $p_\xi\leq p_1$. If we take the same $\eta_0$,
$0<\eta_0<\delta-p_1\leq\delta-p_\xi$ and for every $\xi$ we have
\begin{equation}
\label{1eps-lem-f5}
P\left\{\frac{\mu_m^\xi}{m}>\delta\right\}\leq
P\left\{\left|\frac{\mu_m^\xi}{m}-p_\xi\right|\geq\eta_0\right\}\leq\frac{1}{4m\eta_0^2}
\end{equation}

Due to (\ref{1eps-lem-f4}) and (\ref{1eps-lem-f5}), for every $\varepsilon>0$, if we
take $n>\frac{1}{4\varepsilon\eta_0^2}$, 
we get a system $A_n$ which recognizes the language $L$ with interval 
$(\varepsilon_1,1-\varepsilon_2)$, where $\varepsilon_1,\varepsilon_2<\varepsilon$.

Let us show that $A_n$ can be simulated by a PRA-C. The
automaton $A'=(Q',\Sigma, q'_0, Q'_F, \delta')$ is constructed as follows:

$Q'\stackrel{\scriptscriptstyle {\rm def}}{=}
\left\{\langle q_{s_1}q_{s_2}\dots q_{s_n}\rangle\ |\ 0\leq s_i\leq|Q|-1\right\}$;
$q'_0\stackrel{\scriptscriptstyle {\rm def}}{=}\langle q_0q_0\dots q_0\rangle$.

A sequence $\langle q_{s_1}q_{s_2}\dots q_{s_n}\rangle$ is an accepting state of
$A'$ if more than $n\delta$ elements in the sequence are accepting states of $A$.
We have defined the set $Q'_F$.

Given $\sigma\in\Gamma$,
$\delta'(\langle q_{a_1}q_{a_2}\dots q_{a_n}\rangle,\sigma,\langle q_{b_1}q_{b_2}\dots q_{b_n}\rangle)
\stackrel{\scriptscriptstyle {\rm def}}{=}
\prod\limits_{i=1}^{n}\delta(q_{a_i},\sigma,q_{b_i})$.

In essence, $Q'$ is n-th Cartesian power of $Q$ and the linear space formed by $A'$ is n-th tensor 
power of the linear space formed by $A$.
If we take a symbol $\sigma\in\Gamma$, transition is determined by
$|Q|^n\times|Q|^n$ matrix $V'_\sigma$, which
is n-th matrix direct power of $V_\sigma$, i.e, $V'_\sigma=\bigotimes\limits_{i=1}^{n}V_\sigma$.

$A'$ simulates the system $A_n$. Since matrix direct product of two doubly stochastic matrices 
is a doubly stochastic matrix, $\forall\sigma$ $V'_\sigma$ are doubly stochastic matrices.
Therefore our automaton $A'$ is a PRA-C.

We have proved that $\forall\varepsilon>0$ the language $L$ is recognized by some PRA-C with
interval $(\varepsilon_1,1-\varepsilon_2)$, where $\varepsilon_1,\varepsilon_2<\varepsilon$.
Therefore the language $L$ is recognized with probability $1-\varepsilon$.
\qed
\end{proof}

\begin{lemma}
\label{union-lemma}
If a language $L_1$ is recognizable with probability greater than $\frac{2}{3}$ and
a language $L_2$ is recognizable with probability greater than $\frac{2}{3}$ then
languages $L_1\cap L_2$ and $L_1\cup L_2$ are recognizable with probability greater
than $\frac{1}{2}$.
\end{lemma}
\begin{proof}
Let us consider automata $A=(Q_A,\Sigma,q_{0,A},Q_{F,A},\delta_A)$ and\\
$B=(Q_B,\Sigma,q_{0,B},Q_{F,B},\delta_B)$ which recognize the languages
$L_1$, $L_2$ with probabilities $p_1,p_2>\frac{2}{3}$, respectively.
Let us assume that $A,B$ have $m$ and $n$ states, respectively.
Without loss of generality we can assume that $p_1\leq p_2$.

Informally, having read end-marker symbol $\#$, with probability
$\frac{1}{2}$ we simulate the automaton $A_1$ and with the same
probability we simulate the automaton $A_2$.

Formally, we construct an automaton $C=(Q,\Sigma,q_0,Q_F,\delta)$ with
the following properties.\\
$Q\stackrel{\scriptscriptstyle {\rm def}}{=}Q_A\cup Q_B$;
$q_0\stackrel{\scriptscriptstyle {\rm def}}{=}q_{0,A}$;
$Q_F\stackrel{\scriptscriptstyle {\rm def}}{=}Q_{F,A}\cup Q_{F,B}$;
$\delta\stackrel{\scriptscriptstyle {\rm def}}{=}\delta_A\cup\delta_B$, with
an exception that:
\begin{itemize}
\item $\delta(q_0,\#,q_{i,A})=\frac{1}{2}\delta_A(q_0,\#,q_{i,A})$;
\item $\delta(q_0,\#,q_{i,B})=\frac{1}{2}\delta_B(q_0,\#,q_{i,B})$;
\item $\forall q_i,\ q_i\neq q_0,\ \delta(q_i,\#,q)=\frac{1-\delta(q_0,\#,q)}{m+n-1}$.
\end{itemize}
Since $\delta$ satisfies Definition \ref{prac-def}, our construction of PRA-C is complete.

The automaton $C$ recognizes the language $L_1\cap L_2$ with interval 
$(p,\frac{p_1+p_2}{2})$, where $p\leq 1-\frac{1}{2}p_1$. (Since $p_1,p_2>\frac{2}{3}$, 
$1-\frac{1}{2}p_1<\frac{p_1+p_2}{2}$)

The automaton $C$ recognizes the language $L_1\cup L_2$ with interval
$(\frac{2-p_1-p_2}{2},p)$, where $p\geq\frac{1}{2}p_1$.
(Again, $\frac{2-p_1-p_2}{2}<\frac{1}{2}p_1$)

Therefore by Lemma \ref{cpm-lem}, the languages $L_1\cap L_2$ and $L_1\cup L_2$ are
recognizable with probabilities greater than $\frac{1}{2}$.
\qed
\end{proof}

\begin{theorem}
The class of languages recognized by PRA-C is closed under intersection,
union and complement.
\end{theorem}
\begin{proof}
Let us consider languages $L_1,L_2$ recognized by some PRA-C automata.
By Theorem \ref{1eps-lem}, these languages is recognizable with probability
$1-\varepsilon$, and therefore by Lemmas \ref{cpm-lem} and \ref{union-lemma}, 
union and intersection of these languages are also recognizable.
If a language $L$ is recognizable by a PRA-C $A$, we can construct
an automaton which recognizes a language $\overline L$ just by making
accepting states of $A$ to be rejecting, and vice versa.
\qed
\end{proof}

It is natural to ask what are the languages recognized by PRA-C with probability exactly
$1$.

\begin{theorem}
If a language is recognized by a PRA-C with probability $1$, the language is recognized by 
a permutation automaton.
\end{theorem}
\begin{proof}
Let us consider a language $L$ and a PRA-C $A$, which recognizes
$L$ with probability $1$.

If a word is in $L$, the automaton $A$ has to accept the word with
probability $1$. Conversely, if a word is not in $L$, the word must be
accepted with probability $0$. Therefore,
\begin{equation}
\label{prob1}
\forall q\in Q\ \forall\omega\in\Sigma^*\ 
\mathrm{either}\ q\omega\subseteq Q_F\mathrm{,\ or\ }q\omega\subseteq\overline Q_F.
\end{equation}
Consider a relation between the states of $A$ defined as

$R=\{(q_i,q_j)\ |\ \forall\omega\ q_i\omega\subseteq Q_F\Leftrightarrow q_j\omega\subseteq Q_F\}$.
$R$ is symmetric, reflexive and transitive, therefore $Q$ can be partitioned into equivalence classes
$Q/R=\{[q_0],[q_{i_1}],\dots,[q_{i_k}]\}$. Suppose $A$ is in a state $q$. Due to (\ref{prob1}),
$\forall\omega\ \exists n\ q\omega\subseteq[q_{i_n}]$. In fact, having read a symbol in the alphabet,
$A$ goes from one equivalence class to another with probability $1$.

Hence it is possible to construct the following deterministic automaton $D$, which simulates
$A$. The states are $s_0,\dots,s_k$ and $s_n\sigma=s_m$ iff $[q_{i_n}]\sigma\subseteq[q_{i_m}]$ and
$s_n$ is an accepting state iff $[q_{i_n}]\subseteq Q_F$.
Since all transition matrices of $A$ are doubly stochastic, all transition matrices
of $D$ are permutation matrices.
\qed
\end{proof}

\begin{theorem}
The class of languages recognized by PRA-C is closed under inverse homomorphisms.
\end{theorem}
\begin{proof}
Let us consider finite alphabets $\Sigma, T$, a homomorphism $h:\Sigma\longrightarrow T^*$,
a language $L\subseteq T^*$ and a PRA-C $A=(Q,T,q_0,Q_F,\delta)$, which recognizes $L$ with
interval $(p_1,p_2)$. We prove that exists an automaton $B=(Q,\Sigma,q_0,Q_F,\delta')$ which
recognizes the language $h^{-1}(L)$.

Transition function $\delta$ of A sets transition matrices $V_\tau$, where $\tau\in T$.
To determine $\delta'$, we define transition matrices $V_\sigma$, $\sigma\in\Sigma$.
Let us define a transition matrix $V_{\sigma_k}$:
$$V_{\sigma_k}=V_{[h(\sigma_k)]_m}V_{[h(\sigma_k)]_{m-1}}\dots V_{[h(\sigma_k)]_1},$$
where $m=|h(\sigma_k)|$.
Multiplication of two doubly stochastic matrices is a doubly stochastic matrix, therefore
$B$ is a PRA-C.
Automaton $B$ recognizes $h^{-1}(L)$ with the same interval $(p_1,p_2)$.
\qed
\end{proof}

\begin{corollary}
The class of languages recognized by PRA-C is closed under word quotient.
\end{corollary}
\begin{proof}
This follows from closure under inverse homomorphisms and presence of end-markers $\#,\$$.
\qed
\end{proof}

Even if PRA-C without end-markers are considered, closure under word quotient remains true.
See Appendix for details.

\begin{lemma}
\label{unif-lemma}
If $A$ is a doubly stochastic matrix and $X$ - a vector, then\\
\mbox{$\max(X)\geq\max(AX)$} and $\min(X)\leq\min(AX)$.
\end{lemma}
\begin{proof}
Let us consider
$
X=\left(
\begin{array}{c}
x_1\\
x_2\\
\dots\\
x_n\\
\end{array}
\right)
$
and
$
A=\left(
\begin{array}{cccc}
a_{11} & a_{12} & \dots & a_{1n}\\
a_{21} & a_{22} & \dots & a_{2n}\\
\dots & \dots & \dots & \dots\\
a_{n1} & a_{n2} & \dots & a_{nn}\\
\end{array}
\right)
$,
where $A$ is doubly stochastic.
Let us suppose that $x_j=\max(X)$. For any $i$, $1\leq i\leq n$,
$$
x_j=a_{i1}x_j+a_{i2}x_j+\dots+a_{in}x_j\geq a_{i1}x_1+a_{i2}x_2+\dots+a_{in}x_n.
$$
Therefore $x_j$ is greater or equal than any component of $AX$. The second inequality is
proved in the same way.
\qed
\end{proof}

\begin{theorem}
\label{t-cl-rec}
For every natural positive $n$, a language $L_n=a_1^*a_2^*\dots a_n^*$ is recognizable
by some PRA-C with alphabet $\{a_1,a_2,\dots,a_n\}$.
\end{theorem}
\begin{proof}
We construct a PRA-C with $n+1$ states, $q_0$ being the initial state, corresponding to
probability distribution vector
$
\left(
\begin{array}{c}
1\\
0\\
\dots\\
0\\
\end{array}
\right)
$.
The transition function is determined by $(n+1)\times(n+1)$ matrices\\
$
V_{a_1}=\left(
\begin{array}{ccccc}
1 & 0 & \dots & 0\\
0 & \frac{1}{n} & \dots & \frac{1}{n}\\
\vdots & \vdots & \ddots & \vdots\\
0 & \frac{1}{n} & \dots & \frac{1}{n}\\
\end{array}
\right)
$,
$
V_{a_2}=\left(
\begin{array}{ccccc}
\frac{1}{2} & \frac{1}{2} & 0 & \dots & 0\\
\frac{1}{2} & \frac{1}{2} & 0 & \dots & 0\\
0 & 0 & \frac{1}{n-1} & \dots & \frac{1}{n-1}\\
\vdots & \vdots & \vdots & \ddots & \vdots\\
0 & 0 & \frac{1}{n-1} & \dots & \frac{1}{n-1}\\
\end{array}
\right)
$,
\dots,
$
V_{a_n}=\left(
\begin{array}{ccccc}
\frac{1}{n} & \dots & \frac{1}{n} & 0\\
\vdots & \ddots & \vdots & \vdots\\
\frac{1}{n} & \dots & \frac{1}{n} & 0\\
0 & \dots & 0 & 1\\
\end{array}
\right)
$.
The accepting states are $q_0\dots q_{n-1}$, the only rejecting state is $q_n$.
We prove, that the automaton recognizes the language $L_n$.

Case $\omega\in L_n$.
Having read $\omega\in a_1^*\ldots a_{k-1}^*a_k^+$, the automaton is in
probability distribution
$
\left(
\begin{array}{c}
\frac{1}{k}\\
\dots\\
\frac{1}{k}\\
0\\
\dots\\
0\\
\end{array}
\right)
$.
Therefore all $\omega\in L_n$ are accepted with probability 1.

Case $\omega\notin L_n$.
Consider $k$ such that $\omega=\omega_1\sigma\omega_2$, $|\omega_1|=k$,
$\omega_1\in L_n$ and $\omega_1\sigma\notin L_n$. Since all one-letter words are
in $L_n$, $k>0$. Let $a_t=[\omega]_{k}$ and $a_s=\sigma$.
So we have $s<t$, $1\leq s\leq n-1$, $2\leq t\leq n$.
Having read $\omega_1\in a_1^*\ldots a_{t-1}^*a_t^+$, the automaton is in
distribution
$
\left(
\begin{array}{c}
\frac{1}{t}\\
\dots\\
\frac{1}{t}\\
0\\
\dots\\
0\\
\end{array}
\right)
$.
After that, having read $a_s$, the automaton is in distribution
$
\left(
\begin{array}{cccccc}
\frac{1}{s} & \dots & \frac{1}{s} & 0 & \dots & 0\\
\dots & \dots & \dots & \dots & \dots & \dots\\
\frac{1}{s} & \dots & \frac{1}{s} & 0 & \dots & 0\\
0 & \dots & 0 & \frac{1}{n-s+1} & \dots & \frac{1}{n-s+1}\\
\dots & \dots & \dots & \dots & \dots & \dots\\
0 & \dots & 0 & \frac{1}{n-s+1} & \dots & \frac{1}{n-s+1}\\
\end{array}
\right)
\left(
\begin{array}{c}
\frac{1}{t}\\
\dots\\
\frac{1}{t}\\
0\\
\dots\\
0\\
\end{array}
\right)
=
\left(
\begin{array}{l}
\left.
\begin{array}{c}
\ \ \ \ \frac{1}{t}\ \ \ \\
\ \ \ \ \dots\ \ \ \\
\ \ \ \ \frac{1}{t}\ \ \ \\
\end{array}
\right\}\scriptstyle{s}\\
\left.
\begin{array}{c}
\frac{t-s}{t(n-s+1)}\\
\dots\\
\frac{t-s}{t(n-s+1)}\\
\end{array}
\right\}\scriptstyle{n-s+1}\\
\end{array}
\right).
$
So the word $\omega_1 a_s$ is accepted with probability
$1-\frac{t-s}{t(n-s+1)}$. By Lemma \ref{unif-lemma}, since
$\frac{t-s}{t(n-s+1)}<\frac{1}{t}$, reading the symbols succeeding $\omega_1 a_s$
does not increase accepting probability.
Therefore, to find maximum accepting probability for words
not in $L_n$, we have to maximize $1-\frac{t-s}{t(n-s+1)}$,
where $s<t$, $1\leq s\leq n-1$, $2\leq t\leq n$. Solving this problem, 
we get $t=k+1,s=k$ for $n=2k$, and we get
$t=k+1,s=k$ or $t=k+2,s=k+1$ for $n=2k+1$. So the
maximum accepting probability is $1-\frac{1}{(k+1)^2}$, if $n=2k$, 
and it is $1-\frac{1}{(k+1)(k+2)}$, if $n=2k+1$. All in all,
the automaton recognizes the language with interval
$\left(1-\frac{1}{\left\lfloor(\frac{n}{2})^2\right\rfloor+n+1},\ \ 1\right)$.
(Actually, by Theorem \ref{1eps-lem}, $L_n$ can be recognized with probability
$1-\varepsilon$).
\qed
\end{proof}

\begin{corollary}
Quantum finite automata with mixed states (model of Nayak, \cite{N 99}) recognize
$L_n=a_1^*a_2^*\dots a_n^*$ with probability $1-\varepsilon$.
\end{corollary}
\begin{proof}
This comes from the fact, that matrices $V_{a_1},V_{a_2},\dots,V_{a_n}$ 
from the proof of Theorem \ref{t-cl-rec} (as well as
tensor powers of those matrices) all have
unitary prototypes (see Definition \ref{proto}).
\qed
\end{proof}

\begin{definition}
We say that a regular language is of type $(*)$ if the 
following is true for the minimal deterministic automaton
recognizing this language:
Exist three states $q$, $q_1$, $q_2$, exist words $x$, $y$ such that
\begin{enumerate}
\item $q_1\neq q_2$;
\item $qx=q_1$, $qy=q_2$;
\item $q_1x=q_1$, $q_2y=q_2$;
\item $\forall t\in(x,y)^*\ \exists t_1\in(x,y)^*\ q_1tt_1=q_1$;
\item $\forall t\in(x,y)^*\ \exists t_2\in(x,y)^*\ q_2tt_2=q_2$.
\end{enumerate}
\end{definition}
%
%***************************************************
%\begin{figure}[tb]
%  \centering
%  \begin{Large}
%
%  \include{typ}
% \end{Large}
%  \caption{Type $(*)$ construction}
%  \label{type}
%\end{figure}
%***************************************************
%
%\pagebreak
\begin{definition}
We say that a regular language is of type $(*^\prime)$ if the 
following is true for the minimal deterministic automaton
recognizing this language:
Exist three states $q$, $q_1$, $q_2$, exist words $x$, $y$ such that
%\newpage
\begin{enumerate}
\item $q_1\neq q_2$;
\item $qx=q_1$, $qy=q_2$;
\item $q_1x=q_1$, $q_1y=q_1$;
\item $q_2x=q_2$, $q_2y=q_2$.
\end{enumerate}
\end{definition}
%
%***************************************************
\begin{figure}[htb]
\begin{minipage}{0.4\textwidth}
  \centering
  \begin{Large}
  %TexCad Options
%\grade{\on}
%\emlines{\on}
%\beziermacro{\off}
%\reduce{\on}
%\snapping{\off}
%\quality{2.00}
%\graddiff{0.01}
%\snapasp{1}
%\zoom{1.00}
\special{em:linewidth 0.4pt}
\unitlength 1.00mm
\linethickness{0.4pt}
\begin{picture}(53.00,24.00)
\put(27.33,19.00){\circle{10.00}}
\put(37.33,-1.00){\circle{10.00}}
\put(17.33,-1.00){\circle{10.00}}
%\bezvec{52}(30.66,15.33)(35.00,11.00)(36.66,4.00)
%\put(36.66,4.00){\vector(1,-3){1.18}}
\put(35.75,7.09){\vector(1,-3){1.18}}
\emline{30.66}{15.33}{1}{32.23}{13.57}{2}
\emline{32.23}{13.57}{3}{33.61}{11.60}{4}
\emline{33.61}{11.60}{5}{34.78}{9.45}{6}
\emline{34.78}{9.45}{7}{35.75}{7.09}{8}
%\emline{35.75}{7.09}{9}{36.66}{4.00}{10}
%\end
%\bezvec{52}(24.33,15.00)(19.33,11.00)(18.66,4.00)
%\put(18.66,4.00){\vector(-1,-4){0.88}}
\put(19.30,7.07){\vector(-1,-4){0.88}}
\emline{24.33}{15.00}{11}{22.57}{13.35}{12}
\emline{22.57}{13.35}{13}{21.13}{11.48}{14}
\emline{21.13}{11.48}{15}{20.01}{9.39}{16}
\emline{20.01}{9.39}{17}{19.20}{7.07}{18}
%\emline{19.20}{7.07}{19}{18.66}{4.00}{20}
%\end
%\bezvec{44}(42.00,1.33)(47.33,2.67)(53.00,1.33)
%\put(53.00,1.33){\vector(4,-1){3.52}}
\put(49.43,1.91){\vector(4,-1){3.52}}
\emline{42.00}{1.33}{21}{44.44}{1.80}{22}
\emline{44.44}{1.80}{23}{46.92}{1.99}{24}
\emline{46.92}{1.99}{25}{49.43}{1.91}{26}
%\emline{49.43}{1.91}{27}{53.00}{1.33}{28}
%\end
%\bezvec{48}(53.00,-3.00)(47.33,-4.33)(41.66,-3.00)
%\put(41.66,-3.00){\vector(-4,1){3.52}}
\put(45.92,-3.62){\vector(-4,1){3.52}}
\emline{53.00}{-3.00}{29}{50.64}{-3.44}{30}
\emline{50.64}{-3.44}{31}{48.28}{-3.65}{32}
\emline{48.28}{-3.65}{33}{45.92}{-3.62}{34}
%\emline{45.92}{-3.62}{35}{41.66}{-3.00}{36}
%\end
%\bezvec{48}(13.00,1.00)(8.66,2.67)(1.66,1.67)
\put(1.66,1.67){\vector(-1,0){0.2}}
\emline{13.00}{1.00}{37}{11.08}{1.58}{38}
\emline{11.08}{1.58}{39}{8.92}{1.93}{40}
\emline{8.92}{1.93}{41}{1.66}{1.67}{42}
%\end
%\bezvec{48}(1.66,-3.00)(6.66,-4.33)(13.00,-3.33)
\put(13.00,-3.33){\vector(1,0){0.2}}
\emline{1.66}{-3.00}{43}{3.80}{-3.45}{44}
\emline{3.80}{-3.45}{45}{6.06}{-3.70}{46}
\emline{6.06}{-3.70}{47}{8.44}{-3.75}{48}
\emline{8.44}{-3.75}{49}{13.00}{-3.33}{50}
%\end
\put(27.33,19.00){\makebox(0,0)[cc]{$q$}}
\put(17.33,-1.00){\makebox(0,0)[cc]{$q_1$}}
\put(37.33,-1.00){\makebox(0,0)[cc]{$q_2$}}
\put(19.33,12.00){\makebox(0,0)[cc]{$x$}}
\put(35.33,12.00){\makebox(0,0)[cc]{$y$}}
\put(5.66,3.50){\makebox(0,0)[cc]{$t$}}
\put(47.00,3.50){\makebox(0,0)[cc]{$t$}}
\put(5.66,-1.67){\makebox(0,0)[cc]{$t_1$}}
\put(47.00,-1.67){\makebox(0,0)[cc]{$t_2$}}
\put(41.33,2.00){\vector(-1,-2){0.2}}
\emline{38.00}{3.67}{51}{39.48}{5.39}{52}
\emline{39.48}{5.39}{53}{40.71}{6.71}{54}
\emline{40.71}{6.71}{55}{41.70}{7.63}{56}
\emline{41.70}{7.63}{57}{42.44}{8.15}{58}
\emline{42.44}{8.15}{59}{42.93}{8.27}{60}
\emline{42.93}{8.27}{61}{43.18}{7.99}{62}
\emline{43.18}{7.99}{63}{43.18}{7.31}{64}
\emline{43.18}{7.31}{65}{42.93}{6.23}{66}
\emline{42.93}{6.23}{67}{42.44}{4.75}{68}
\emline{42.44}{4.75}{69}{41.33}{2.00}{70}
%\end
%\bezvec{100}(16.67,3.67)(9.00,13.67)(13.33,2.00)
\put(13.33,2.00){\vector(1,-2){0.2}}
\emline{16.67}{3.67}{71}{15.25}{5.45}{72}
\emline{15.25}{5.45}{73}{14.08}{6.80}{74}
\emline{14.08}{6.80}{75}{13.15}{7.72}{76}
\emline{13.15}{7.72}{77}{12.45}{8.20}{78}
\emline{12.45}{8.20}{79}{12.00}{8.25}{80}
\emline{12.00}{8.25}{81}{11.79}{7.87}{82}
\emline{11.79}{7.87}{83}{11.81}{7.05}{84}
\emline{11.81}{7.05}{85}{12.08}{5.80}{86}
\emline{12.08}{5.80}{87}{12.59}{4.12}{88}
\emline{12.59}{4.12}{89}{13.33}{2.00}{90}
%\end
\put(10.00,9.00){\makebox(0,0)[cc]{$x$}}
\put(45.00,9.00){\makebox(0,0)[cc]{$y$}}
\end{picture}
  \end{Large}
  \caption{Type $(*)$ construction}
  \label{type}
\end{minipage}
\hspace{0.1\textwidth}
\begin{minipage}{0.4\textwidth}
  \centering
  \begin{Large}
%TexCad Options
%\grade{\on}
%\emlines{\on}
%\beziermacro{\off}
%\reduce{\on}
%\snapping{\off}
%\quality{2.00}
%\graddiff{0.01}
%\snapasp{1}
%\zoom{1.00}
\special{em:linewidth 0.4pt}
\unitlength 1.00mm
\linethickness{0.4pt}
\begin{picture}(52.67,24.00)
\put(26.00,19.00){\circle{10.00}}
\put(36.00,-1.00){\circle{10.00}}
\put(16.00,-1.00){\circle{10.00}}
%\bezvec{52}(29.33,15.33)(33.67,11.00)(35.33,4.00)
\put(34.42,7.09){\vector(1,-3){1.18}}
\emline{29.33}{15.33}{1}{30.90}{13.57}{2}
\emline{30.90}{13.57}{3}{32.27}{11.60}{4}
\emline{32.27}{11.60}{5}{33.45}{9.45}{6}
\emline{33.45}{9.45}{7}{34.42}{7.09}{8}
%\emline{34.42}{7.09}{9}{35.33}{4.00}{10}
%\end
%\bezvec{52}(23.00,15.00)(18.00,11.00)(17.33,4.00)
\put(17.87,7.07){\vector(-1,-4){0.88}}
\emline{23.00}{15.00}{11}{21.24}{13.35}{12}
\emline{21.24}{13.35}{13}{19.79}{11.48}{14}
\emline{19.79}{11.48}{15}{18.67}{9.39}{16}
\emline{18.67}{9.39}{17}{17.87}{7.07}{18}
%\emline{17.87}{7.07}{19}{17.33}{4.00}{20}
%\end
\put(26.00,19.00){\makebox(0,0)[cc]{$q$}}
\put(16.00,-1.00){\makebox(0,0)[cc]{$q_1$}}
\put(36.00,-1.00){\makebox(0,0)[cc]{$q_2$}}
\put(18.00,12.00){\makebox(0,0)[cc]{$x$}}
\put(34.00,12.00){\makebox(0,0)[cc]{$y$}}
%\bezvec{104}(40.00,2.00)(52.67,-1.67)(40.33,-4.00)
\put(43.20,-3.35){\vector(-4,-1){3.52}}
\emline{40.00}{2.00}{21}{42.21}{1.31}{22}
\emline{42.21}{1.31}{23}{43.95}{0.64}{24}
\emline{43.95}{0.64}{25}{45.23}{-0.01}{26}
\emline{45.23}{-0.01}{27}{46.05}{-0.62}{28}
\emline{46.05}{-0.62}{29}{46.40}{-1.22}{30}
\emline{46.40}{-1.22}{31}{46.29}{-1.79}{32}
\emline{46.29}{-1.79}{33}{45.73}{-2.33}{34}
\emline{45.73}{-2.33}{35}{44.69}{-2.85}{36}
\emline{44.69}{-2.85}{37}{43.20}{-3.35}{38}
%\emline{43.20}{-3.35}{39}{40.33}{-4.00}{40}
%\end
%\bezvec{104}(12.00,2.00)(-0.67,-1.67)(12.00,-4.00)
\put(9.05,-3.35){\vector(4,-1){3.52}}
\emline{12.00}{2.00}{41}{9.80}{1.31}{42}
\emline{9.80}{1.31}{43}{8.06}{0.64}{44}
\emline{8.06}{0.64}{45}{6.80}{-0.01}{46}
\emline{6.80}{-0.01}{47}{6.00}{-0.62}{48}
\emline{6.00}{-0.62}{49}{5.67}{-1.22}{50}
\emline{5.67}{-1.22}{51}{5.81}{-1.79}{52}
\emline{5.81}{-1.79}{53}{6.42}{-2.33}{54}
\emline{6.42}{-2.33}{55}{7.50}{-2.85}{56}
\emline{7.50}{-2.85}{57}{9.05}{-3.35}{58}
%\emline{9.05}{-3.35}{59}{12.00}{-4.00}{60}
%\end
\put(50.00,-0.33){\makebox(0,0)[cc]{$x,y$}}
\put(2.33,-0.33){\makebox(0,0)[cc]{$x,y$}}
\end{picture}
  \end{Large}
  \caption{Type $(*^\prime)$ construction}
  \label{type1}
\end{minipage}
\end{figure}
%***************************************************
\newpage
\begin{definition}
We say that a regular language is of type $(*^{\prime\prime})$ if the 
following is true for the minimal deterministic automaton
recognizing this language:
Exist two states $q_1$, $q_2$, exist words $x$, $y$ such that
\begin{enumerate}
\item $q_1\neq q_2$;
\item $q_1x=q_2$, $q_2x=q_2$;
\item $q_2y=q_1$.
\end{enumerate}
\end{definition}
%
%***************************************************
\begin{figure}[htb]
  \centering
  \begin{Large}
%TexCad Options
%\grade{\on}
%\emlines{\on}
%\beziermacro{\off}
%\reduce{\on}
%\snapping{\off}
%\quality{2.00}
%\graddiff{0.01}
%\snapasp{1}
%\zoom{1.00}
\special{em:linewidth 0.4pt}
\unitlength 1.00mm
\linethickness{0.4pt}
\begin{picture}(43,4.00)
\put(25.67,4.00){\circle{10.00}}
\put(5.67,4.00){\circle{10.00}}
\put(5.67,4.00){\makebox(0,0)[cc]{$q_1$}}
\put(25.67,4.00){\makebox(0,0)[cc]{$q_2$}}
%\bezvec{104}(29.67,2.00)(42.34,-1.67)(30.00,-4.00)
\put(32.87,1.65){\vector(-4,-1){3.52}}
\emline{29.67}{7.00}{1}{31.87}{6.31}{2}
\emline{31.87}{6.31}{3}{33.61}{5.64}{4}
\emline{33.61}{5.64}{5}{34.90}{4.99}{6}
\emline{34.90}{4.99}{7}{35.71}{4.38}{8}
\emline{35.71}{4.38}{9}{36.07}{3.78}{10}
\emline{36.07}{3.78}{11}{35.96}{3.21}{12}
\emline{35.96}{3.21}{13}{35.39}{2.67}{14}
\emline{35.39}{2.67}{15}{34.36}{2.15}{16}
\emline{34.36}{2.15}{17}{32.87}{1.65}{18}
%\emline{32.87}{1.65}{19}{30.00}{1.00}{20}
%\end
\put(39.67,4.67){\makebox(0,0)[cc]{$x$}}
%\bezvec{48}(9.67,2.00)(15.67,2.67)(21.67,2.00)
\put(21.67,7.00){\vector(1,0){0.2}}
\emline{9.67}{7.00}{21}{12.17}{7.22}{22}
\emline{12.17}{7.22}{23}{14.67}{7.33}{24}
\emline{14.67}{7.33}{25}{17.17}{7.31}{26}
\emline{17.17}{7.31}{27}{21.67}{7.00}{28}
%\end
%\bezvec{48}(21.67,-4.00)(15.67,-4.67)(9.67,-4.00)
\put(9.67,1.00){\vector(-1,0){0.2}}
\emline{21.67}{1.00}{29}{19.17}{0.78}{30}
\emline{19.17}{0.78}{31}{16.67}{0.67}{32}
\emline{16.67}{0.67}{33}{14.17}{0.69}{34}
\emline{14.17}{0.69}{35}{9.67}{1.00}{36}
%\end
\put(15.34,8.67){\makebox(0,0)[cc]{$x$}}
\put(15.00,2.00){\makebox(0,0)[cc]{$y$}}
\end{picture}
  \end{Large}
  \caption{Type $(*^{\prime\prime})$ construction}
  \label{type2}
\end{figure}
%***************************************************

Type $(*^{\prime\prime})$ languages are exactly those languages that violate the partial order condition
of \cite{BP 99}.

\begin{lemma}
\label{ret-lemma}
If $A$ is a deterministic finite automaton with a set of states $Q$ and alphabet $\Sigma$, then
$\forall q\in Q$ $\forall x\in\Sigma^*$ $\exists k>0$ $qx^k=qx^{2k}$.
\end{lemma}
\begin{proof}
We paraphrase a result from the theory of finite semigroups.
Consider a state $q$ and a word $x$. Since number of states is finite,
$\exists m\geq0$ $\exists s\geq1$ $\forall n$ \mbox{$qx^m=qx^mx^{sn}$}.
Take $n_0$, such that $sn_0>m$. Note that $\forall t\geq0$ $qx^{m+t}=qx^{m+t}x^{sn_0}$.
We take $t=sn_0-m$, so $qx^{sn_0}=qx^{sn_0}x^{sn_0}$. Take $k=sn_0$.
\qed
\end{proof}

\begin{lemma}
\label{eqtype-lemma}
A regular language is of type $(*)$ iff it is of type $(*^\prime)$ or type $(*^{\prime\prime})$.
\end{lemma}
\begin{proof}
1) If a language is of type $(*^\prime)$, it is of type $(*)$. Obvious.

2) If a language is of type $(*^{\prime\prime})$, it is of type $(*)$. Consider
a language of type $(*^{\prime\prime})$ with states $q_1^{\prime\prime}, q_2^{\prime\prime}$ and words
$x^{\prime\prime}, y^{\prime\prime}$. To build construction of type $(*)$, we take $q=q_1=q_1^{\prime\prime}$,
$q_2=q_2^{\prime\prime}$, $x=x^{\prime\prime}y^{\prime\prime}$, $y=x^{\prime\prime}$. That forms transitions
$qx=q_1$, $qy=q_2$, $q_1x=q_1$, $q_1y=q_2$, $q_2x=q_1$, $q_2y=q_2$. We have satisfied all the rules of $(*)$.

3) If a language is of type $(*)$, it is of type $(*^\prime)$ or $(*^{\prime\prime})$.
Consider a language whose minimal deterministic automaton has construction $(*)$.
By Lemma \ref{ret-lemma},
\begin{verse}
$\exists t\exists b$ $q_1y^b=q_t$ and $q_ty^b=q_t$;\\
$\exists u\exists c$ $q_2x^c=q_u$ and $q_ux^c=q_u$.\\
\end{verse}
If $q_1\neq q_t$, by the 4th rule of $(*)$, $\exists z\ q_tz=q_1$. Therefore the language is
of type $(*^{\prime\prime})$.
If $q_2\neq q_u$, by the 5th rule of $(*)$, $\exists z\ q_uz=q_2$, and the language is
of type $(*^{\prime\prime})$.\\
If $q_1=q_t$ and $q_2=q_u$, we have $qx^c=q_1$, $qy^b=q_2$, $q_1x^c=q_1y^b=q_1$,
$q_2x^c=q_2y^b=q_2$. We get the construction $(*^\prime)$ if we take $x^\prime=x^c$, $y^\prime=y^b$.
\qed
\end{proof}

We are going to prove that every language of type $(*)$ is not recognizable by any PRA-C.
For this purpose, we recall several definitions from the theory of finite Markov chains
(\cite{KS 76}, etc.)

A Markov chain with $n$ states can be determined by an $n\times n$ stochastic matrix $A$, i.e., matrix, where the sum of
elements of every column in the matrix is $1$. If $A_{i,j}=p>0$, it means that a state $q_i$ is accessible from a state
$q_j$ with a positive probability $p$ in one step. Generally speaking, the matrix
depends on the numbering of the states; if the states are renumbered, the matrix changes, as its rows and
columns also need to be renumbered.

\begin{definition}
A state $q_j$ is accessible from $q_i$ (denoted $q_i\rightarrow q_j$) if there
is a positive probability to get from $q_i$ to $q_j$ (possibly in several steps).
\end{definition}

\begin{definition}
States $q_i$ and $q_j$ communicate (denoted $q_i\leftrightarrow q_j$) if $q_i\rightarrow q_j$ and
$q_j\rightarrow q_i$.
\end{definition}

\begin{definition}
A state $q$ is called ergodic if $\forall i\ q\rightarrow q_i\Rightarrow q_i\rightarrow q$.
Otherwise the state is called transient.
\end{definition}

\begin{definition}
A Markov chain without transient states is called irreducible if
for all $q_i,q_j$ $q_i\leftrightarrow q_j$. Otherwise the chain without transient states is called reducible.
\end{definition}

\begin{definition}
The period of an ergodic state $q_i\in Q$ of a Markov chain with a matrix $A$ is defined as
$d(q_i)=\gcd\{n>0\ |\ (A^n)_{i,i}>0\}$.
\end{definition}

\begin{definition}
\label{st-per-def}
An ergodic state $q_i$ is called aperiodic if $d(q_i)=1$. Otherwise the ergodic state is called periodic.
\end{definition}

\begin{definition}
A Markov chain without transient states is called aperiodic if all its states are aperiodic.
Otherwise the chain without transient states is called periodic.
\end{definition}

\begin{definition}
A probability distribution X of a Markov chain with a matrix $A$ is called stationary, if $AX=X$.
\end{definition}

\begin{definition}
A Markov chain is called doubly stochastic, if its transition matrix is a doubly stochastic matrix.
\end{definition}

We recall the following theorem from the theory of finite Markov chains:
\begin{theorem}
\label{therstat}
If a Markov chain with a matrix $A$ is irreducible and aperiodic, then\\
a) it has a unique stationary distribution $Z$;\\
b) $\lim\limits_{n\to\infty}A^n=(Z,\dots,Z)$;\\
c) $\forall X$ $\lim\limits_{n\rightarrow\infty}A^nX=Z$.
\end{theorem}

\begin{corollary}
\label{corstat}
If a doubly stochastic Markov chain with an $m\times m$ matrix A is irreducible and aperiodic,\\
a) $\lim\limits_{n\to\infty}A^n=
\left(
\begin{array}{ccc}
\frac{1}{m} & \dots & \frac{1}{m}\\
\dots & \dots & \dots\\
\frac{1}{m} & \dots & \frac{1}{m}\\
\end{array}
\right)
$;\\
b) $\forall X$ $\lim\limits_{n\rightarrow\infty}A^nX=
\left(
\begin{array}{c}
\frac{1}{m}\\
\dots\\
\frac{1}{m}\\
\end{array}
\right)
$.
\end{corollary}
\begin{proof}
By Theorem \ref{therstat}.
\qed
\end{proof}

\begin{lemma}
\label{reflex}
If $M$ is a doubly stochastic Markov chain with a matrix $A$, then $\forall q$ $q\rightarrow q$.
\end{lemma}
\begin{proof}
Assume existence of $q_0$ such that $q_0$ is not accessible from itself.
Let $Q_{q_0}=\{q_i\ |\ q_0\rightarrow q_i\}=\{q_1,\dots,q_k\}$.
$Q_{q_0}$ is not empty set.
Consider those rows and columns of $A$,
which are indexed by states in $Q_{q_0}$. These rows and columns form a submatrix $A^\prime$.
Each column $j$ of $A^\prime$ must include all non-zero elements of the corresponding column of
$A$ as those states are accessible from the state $q_j$, hence also from $q_0$ and are in $Q_{q_0}$.
Therefore $\forall j$, $1\leq j\leq k$, $\sum\limits_{i=1}^k A_{i,j}^\prime=1$ and 
$\sum\limits_{1\leq i,j\leq k} A_{i,j}^\prime=k$. On the other hand, since $q_0\notin Q_{q_0}$, 
a row of $A^\prime$ indexed by a state accessible in one step from $q_0$ does not include all nonzero
elements. Since A is doubly stochastic, $\exists i, 1\leq i\leq k$, $\sum\limits_{j=1}^k A^\prime_{i,j}<1$
and $\sum\limits_{1\leq i,j\leq k} A_{i,j}^\prime<k$. This is a contradiction.
\qed
\end{proof}

\begin{corollary}
\label{ppow}
Suppose $A$ is a doubly stochastic matrix. Then exists $k>0$, such that
$\forall i$ $(A^k)_{i,i}>0$.
\end{corollary}
\begin{proof}
Consider an $m\times m$ doubly stochastic matrix $A$.
By Lemma \ref{reflex}, $\forall i$ $\exists n_i>0$ $(A^{n_i})_{i,i}>0$. Take $n=\prod\limits_{s=1}^m n_s$.
For every $i$, $(A^{n})_{i,i}>0$.
\qed
\end{proof}

\begin{lemma}
\label{quasi-sim-lem}
If $M$ is a doubly stochastic Markov chain with a matrix $A$, then $\forall q_a,q_b$
$A_{b,a}>0\Rightarrow q_b\rightarrow q_a$.
\end{lemma}
\begin{proof}
$A_{b,a}>0$ means that $q_b$ is accessible from $q_a$ in one step. We have to prove, that $q_b\rightarrow q_a$.
Assume from the contrary, that $q_a$ is not accessible from $q_b$. Let 
$Q_{q_b}=\{q_i\ |\ q_b\rightarrow q_i\}=\{q_1,q_2,\dots,q_k\}$. By Lemma \ref{reflex}, $q_b\in Q_{q_b}$.
As in proof of Lemma \ref{reflex}, consider a matrix $A^\prime$, which is a submatrix of $A$ and whose rows
and columns are indexed by states in $Q_{q_b}$. Each column $j$ has to include all nonzero elements of the 
corresponding column of $A$. Therefore $\forall j$, $1\leq j\leq k$, $\sum\limits_{i=1}^k A_{i,j}^\prime=1$ and
$\sum\limits_{1\leq i,j\leq k} A_{i,j}^\prime=k$. On the other hand, $A_{b,a}>0$ and $q_a\notin Q_{q_b}$, therefore
a row of $A^\prime$ indexed by $q_b$ does not include all nonzero elements.
Since $A$ is doubly stochastic, $\sum\limits_{j=1}^k A^\prime_{b,j}<1$ and
$\sum\limits_{1\leq i,j\leq k} A_{i,j}^\prime<k$. This is a contradiction.
\qed
\end{proof}

\begin{corollary}
\label{cor2}
If $M$ is a doubly stochastic Markov chain and $q_a\rightarrow q_b$, then $q_a\leftrightarrow q_b$.
\end{corollary}
\begin{proof}
If $q_a\rightarrow q_b$ then exists a sequence $q_{i_1},q_{i_2},\dots,q_{i_k}$, such that
$A_{i_1,a}>0,\ A_{i_2,i_1}>0,\dots,A_{i_k,i_{k-1}}>0,\ A_{b,i_k}>0$.
By Lemma \ref{quasi-sim-lem}, we get $q_b\rightarrow q_{i_k}$, $q_{i_k}\rightarrow q_{i_{k-1}}$,
$\dots$, $q_{i_2}\rightarrow q_{i_1}$, $q_{i_1}\rightarrow q_a$. Therefore $q_b\rightarrow q_a$.
\qed
\end{proof}

By Corollary \ref{cor2}, every doubly stochastic Markov chain does not have transient states,
so it is either periodic or aperiodic, either reducible or irreducible.
%
%\begin{corollary}
%Suppose $M$ is an aperiodic doubly stochastic Markov chain with a matrix $A$.\\
%Then exists $k>0$, such that $\forall i,j$ $q_i\rightarrow q_j\Rightarrow[(A^k)_{i,j}>0\textrm{ and }(A^k)_{j,i}>0]$.
%\end{corollary}
%\begin{proof}
%By Lemma \ref{reflex}, $\to$ is reflexive. By Corollary \ref{cor2}, $\to$ is symmetric. Surely
%$\to$ is transitive. Therefore all states indexing doubly stochastic matrix may be partitioned into
%equivalence classes (called ergodic sets). Let us renumber states in such a way, that states from one ergodic
%set have consecutive numbers. The resulting matrix is a block diagonal matrix, each block numbered by
%states of an ergodic set. Let us identify these blocks as $A_1,A_2,\dots,A_s$. Each block corresponds to 
%an irreducible aperiodic doubly stochastic Markov chain. Let us consider a $k\times k$ block $A_t$.
%By Corollary \ref{corstat}, $\lim\limits_{n\to\infty}A_t^n=
%\left(
%\begin{array}{ccc}
%\frac{1}{k} & \dots & \frac{1}{k}\\
%\dots & \dots & \dots\\
%\frac{1}{k} & \dots & \frac{1}{k}\\
%\end{array}
%\right)
%$.
%Therefore $\exists N\ \forall i$ all elements of $A_i^N$ are positive. Now
%suppose $q_i\to q_j$. The states $q_i$ and $q_j$ are in one ergodic set, therefore $(A^N)_{i,j}>0$ and
%$(A^N)_{j,i}>0$.
%\qed
%\end{proof}

\begin{lemma}
\label{type2-lemma}
If a regular language is of type $(*^\prime)$, it is not recognizable
by any PRA-C.
\end{lemma}
\begin{proof}
Assume from the contrary, that $A$ is a PRA-C automaton which recognizes a language 
\mbox{$L\subset\Sigma^*$} of type $(*^\prime)$.

Since $L$ is of type $(*^\prime)$, it is recognized by a deterministic automaton $D$ which
has three states $q$, $q_1$, $q_2$ such that 
$q_1\neq q_2$, $qx=q_1$, $qy=q_2$, $q_1x=q_1$, $q_1y=q_1$, $q_2x=q_2$, $q_2y=q_2$, where
$x,y\in\Sigma^*$.
Furthermore, exists $\omega\in\Sigma^*$ such that $q_0\omega=q$, where $q_0$ is an initial state of $D$, and
exists a word $z\in\Sigma^*$, such that $q_1z=q_{acc}$ if and only if $q_2z=q_{rej}$, where $q_{acc}$ is
an accepting state and $q_{rej}$ is a rejecting state of $D$. Without loss
of generality we assume that $q_1z=q_{acc}$ and $q_2z=q_{rej}$.

The transition function of the automaton $A$ is determined by doubly stochastic matrices
$V_{\sigma_1},\dots,V_{\sigma_n}$. The words from the construction $(*^\prime)$ are
$x=\sigma_{i_1}\dots\sigma_{i_k}$ and $y=\sigma_{j_1}\dots\sigma_{j_s}$.
The transitions induced by words $x$ and $y$ are determined by doubly stochastic matrices
$X=V_{\sigma_{i_k}}\!\!\dots V_{\sigma_{i_1}}$ and $Y=V_{\sigma_{j_s}}\!\!\dots V_{\sigma_{j_1}}$.
Similarly, the transitions induced by words $\omega$ and $z$ are determined by doubly stochastic
matrices $W$ and $Z$.
By Corollary \ref{ppow}, exists $K>0$, such that
\begin{equation}
\label{eq-diagpos}
\forall i\ (X^K)_{i,i}>0\textrm{ and }(Y^K)_{i,i}>0.
\end{equation}

Consider a relation between the states of the automaton defined as
$R=\{(q_i,q_j)\ |\ q_i\stackrel{(x^K,y^K)\textrm{*}}{\longrightarrow}q_j\}$.
By (\ref{eq-diagpos}), this relation is reflexive.

Suppose exists a word
$\xi=\xi_1\xi_2\dots\xi_k$, $\xi_s\in\{x^K,y^K\}$, such that $q\stackrel{\xi}{\longrightarrow}q^\prime$.
This means that $q\stackrel{\xi_1}{\longrightarrow}q_{i_1}$,
$q_{i_1}\stackrel{\xi_2}{\longrightarrow}q_{i_2}$,$\dots$, 
$q_{i_{k-1}}\stackrel{\xi_k}{\longrightarrow}q^\prime$.
By Corollary \ref{cor2}, since both $X^K$ and $Y^K$ are doubly stochastic, $\exists\xi^\prime_k\dots\xi^\prime_1$,
$\xi^\prime_s\in\{(x^K)^*,(y^K)^*\}$, such that $q^\prime\stackrel{\xi^\prime_k}{\longrightarrow}q_{i_{k-1}}$,$\dots$,
$q_{i_2}\stackrel{\xi^\prime_2}{\longrightarrow}q_{i_1}$, 
$q_{i_1}\stackrel{\xi^\prime_1}{\longrightarrow}q$, therefore
$q^\prime\stackrel{\xi^\prime}{\longrightarrow}q$, where $\xi^\prime\in(x^K,y^K)^*$.
So the relation $R$ is symmetric.

Surely $R$ is transitive. Therefore all states of $A$ may be partitioned into
equivalence classes $[q_0],[q_{i_1}],\dots,[q_{i_n}]$. Let us renumber the states of $A$
in such a way, that states from one equivalence class have consecutive numbers.
First come the states in $[q_0]$, then in $[q_{i_1}]$, etc.

Consider the word $x^Ky^K$. The transition induced by this word is determined by a doubly stochastic matrix
$C=Y^KX^K$. We prove the following proposition. States $q_a$ and $q_b$ are in one equivalence class if and only if
$q_a\to q_b$ with matrix $C$. Suppose $q_a\to q_b$. Then $(q_a,q_b)\in R$, and $q_a,q_b$ are in one equivalence
class. Suppose $q_a,q_b$ are in one equivalence class. Then 
\begin{equation}
\label{nine}
q_a\stackrel{\xi_1}{\longrightarrow}q_{i_1},
q_{i_1}\stackrel{\xi_2}{\longrightarrow}q_{i_2},\dots, 
q_{i_{k-1}}\stackrel{\xi_k}{\longrightarrow}q_b,\textrm{ where }\xi_s\in\{x^K,y^K\}.
\end{equation}
By (\ref{eq-diagpos}), $q_i\stackrel{x^K}{\longrightarrow}q_i$ and $q_j\stackrel{y^K}{\longrightarrow}q_j$.
Therefore, if $q_i\stackrel{x^K}{\longrightarrow}q_j$, then $q_i\stackrel{x^Ky^K}{\longrightarrow}q_j$, and
again, if $q_i\stackrel{y^K}{\longrightarrow}q_j$, then $q_i\stackrel{x^Ky^K}{\longrightarrow}q_j$.
That transforms (\ref{nine}) to
\begin{equation}
q_a\stackrel{(x^Ky^K)^t}{\longrightarrow}q_b,\textrm{ where }t>0.
\end{equation}
We have proved the proposition.

By the proved proposition, due to the renumbering of states, matrix $C$ is a block diagonal matrix,
where each block corresponds to an equivalence class of the relation $R$. Let us identify these blocks as
$C_0,C_1,\dots,C_n$. By (\ref{eq-diagpos}), a Markov chain with matrix C is aperiodic. Therefore
each block $C_r$ corresponds to an aperiodic irreducible doubly stochastic Markov chain with states $[q_{i_r}]$.
By Corollary \ref{corstat}, $\lim\limits_{m\to\infty}C^m=J$, $J$ is a block diagonal matrix, where for each
$(p\times p)$ block $C_r$ $(C_r)_{i,j}=\frac{1}{p}$.
Relation $q_i\stackrel{(y^K)^*}{\longrightarrow}q_j$ is a subrelation of $R$, therefore $Y^K$ is a block diagonal
matrix with the same block ordering and sizes as $C$ and $J$. (This does not eliminate possibility that 
some block of $Y^K$ is constituted of smaller blocks, however.) Therefore $JY^K=J$, and 
$\lim\limits_{m\to\infty}Z(Y^KX^K)^mW=\lim\limits_{m\to\infty}Z(Y^KX^K)^mY^KW=ZJW$. 
So
\begin{equation}
\forall\varepsilon>0\ \exists m\ \left\|\Big(Z(Y^KX^K)^mW-Z(Y^KX^K)^mY^KW\Big)Q_0\right\|<\varepsilon.
\end{equation}
However, by construction $(*^\prime)$, $\forall k\ \forall m$ $\omega(x^ky^k)^mz\in L$ and
$\omega y^k(x^ky^k)^mz\notin L$. This requires existence of $\varepsilon>0$, such that
\begin{equation}
\forall m\ \left\|\Big(Z(Y^KX^K)^mW-Z(Y^KX^K)^mY^KW\Big)Q_0\right\|>\varepsilon.
\end{equation}
This is a contradiction.
%Schema of the proof.
%
%We take use of Markov chain theory and prove that for words
%$x$, $y$ exists constant $N$, such that for every $\varepsilon$
%exist $D$, such that
%for two non-equivalent words $\omega_1=\nu x^N(y^Nx^N)^D$ and
%$\omega_2=\nu(y^Nx^N)^D$ $|p_{\omega_1}-p_{\omega_2}|<\varepsilon$.
\qed
\end{proof}

\begin{lemma}
\label{type3-lemma}
If a regular language is of type $(*^{\prime\prime})$, it is not recognizable by
any PRA-C.
\end{lemma}
\begin{proof}
Proof is nearly identical to that of Lemma \ref{type2-lemma}.
Consider a PRA-C which recognizes the language $L$ of type $(*^{\prime\prime})$.
We prove that for words $x$, $y$ exists constant $K$, such that
for every $\varepsilon$ exists $m$, such that
for two words $\xi_1=\omega(x^K(xy)^K)^mz$ and
$\xi_2=\omega(x^K(xy)^K)^mx^Kz$, $|p_{\xi_1}-p_{\xi_2}|<\varepsilon$.
We can choose $z$, such that $\xi_1\in L$ iff $\xi_2\notin L$.
\qed
\end{proof}

\begin{theorem}
\label{main-theorem}
If a regular language is of type $(*)$, it is not recognizable by any
PRA-C.
\end{theorem}
\begin{proof}
By Lemmas \ref{eqtype-lemma}, \ref{type2-lemma}, \ref{type3-lemma}.
\qed
\end{proof}

We proved (Lemma \ref{eqtype-lemma}) that the construction of type $(*)$ is a generalization
the construction proposed by \cite{BP 99}. Also it can be easily noticed, that the type $(*)$
construction is a generalization of construction proposed by \cite{AKV 01}.
(Constructions of \cite{BP 99} and \cite{AKV 01} characterize languages, not recognized
by measure-many quantum finite automata of \cite{KW 97}.)

\begin{corollary}
\label{abzv}
Languages (a,b)*a and a(a,b)* are not recognized by PRA-C.
\end{corollary}
\begin{proof}
Both languages are of type $(*)$.
\qed
\end{proof}

\begin{corollary}
Class of languages recognizable by PRA-C is not closed under homomorphisms.
\end{corollary}

\begin{proof}
Consider a homomorphism $a\to a$, $b\to b$, $c\to a$.
Similarly as in Theorem \ref{t-cl-rec}, the language (a,b)*cc* is recognizable by a PRA-C.
(Take $n=2$, $V_a=V_{a_1}$, $V_b=V_{a_1}$, $V_c=V_{a_2}$ from Theorem \ref{t-cl-rec}, $Q_F=\{q_1\}$)
However, by Corollary \ref{abzv} the language (a,b)*aa*=(a,b)*a is not recognizable.
\qed
\end{proof}

\section{1-way Probabilistic Reversible DH-Automata}
\label{Sec1wayDH}
\begin{definition}
The definition differs from one for PRA-C (Definition
\ref{prac-def}) by the following: languages are recognized
according to Definition \ref{dh-acc-def}.
\end{definition}
It is easy to see that the class of languages recognized by PRA-C is a
proper subclass of languages recognized by PRA-DH. For example,
the language a(a,b)* is recognizable by PRA-DH. However, the
following theorem holds:
\begin{theorem}
Language (a,b)*a is not recognized by PRA-DH.
\end{theorem}
\begin{proof}
Assume from the contrary that such automaton exists. While reading
any sequence of a and b, this automaton can halt only with some
probability p strictly less then 1, so accepting and
rejecting probabilities may differ only by 1-p, because any word
belonging to the language is not dependent on any prefix.
Therefore for each $\varepsilon>0$ we can find that after reading
of a prefix of certain length, the total probability to halt while
continue reading the word is less then $\varepsilon$. In this case
we can apply similar techniques as in the proof of Lemma
\ref{type2-lemma}, such that for words $x$, $y$ exists constant
$K$, such that for every $\varepsilon$ exists $s$, such that for
two words $\xi_1=\omega(x^K(xy)^K)^sz$ and
$\xi_2=\omega(x^K(xy)^K)^sx^Kz$,
$|p_{\xi_1}-p_{\xi_2}|<\varepsilon$.
\qed
\end{proof}

\section{Alternative Approach to Finite Reversible
Automata and 1.5-way Probabilistic Reversible Automata}
\label{Sec15way}

Let us consider automaton $A'=(Q,\Sigma,q_0,Q_F,\delta ')$ that
can be obtained from a probabilistic automaton
$A=(Q,\Sigma,q_0,Q_F,\delta)$ by specifying\\
$\delta '
(q,\sigma,q')=\delta(q',\sigma,q)$ for all $q'$, $\sigma$ and $q$.

If $A'$ is a valid probabilistic automaton then we can call $A$
and $A'$ probabilistic reversible automata.

%\textbf{Could be smith as follows ? }

\begin{definition}
An automaton of some type is called \textbf{weakly reversible} if
the reverse of its transition function corresponds to the
transition function of a valid automaton of the same type.

\end{definition}

Note: in case of deterministic automaton where
$\delta:Q\times\Gamma\times Q\longrightarrow \{0,1\} $ this
property means that A' is still deterministic automaton, not
nondeterministic.

In case of one-way automata it is easy to check that this
definition is equivalent to the one in Section \ref{Sec1wayC}.

We give an example that illustrates that in case of 1.5-way
automata these definitions are different.

\begin{definition}
\label{15prc-def1} 1.5-way probabilistic \textbf{weakly}
reversible C-automaton \\
$A=(Q,\Sigma,q_0,Q_F,\delta)$ is
specified by $Q$, $\Sigma$, $q_0$, $Q_F$ defined as in 1-way PRA-C
Definition \ref{prac-def}, and a transition function $$
\delta:Q\times\Gamma\times Q\times D \longrightarrow\bbbr_{[0,1]},
$$ where $\Gamma$ defined as in 1-way PRA-C definition and
$D=\{0,1\}$ denotes whether automaton stays on the same position
or moves one letter ahead on the input tape. Furthermore,
transition function satisfies the following requirements:

\begin{eqnarray}
&&\forall(q_1,\sigma_1)\in Q\times\Gamma\ \sum\limits_{q\in Q,d\in
D}\delta(q_1,\sigma_1,q,d)=1 \label{sc3-1-1}\\
&&\forall(q_1,\sigma_1)\in Q\times\Gamma\ \sum\limits_{q\in Q,d\in
D}\delta(q,\sigma_1,q_1,d)=1 \label{sc3-2-1}
\end{eqnarray}

\end{definition}

\begin{definition}
\label{15prc-def2} 1.5-way probabilistic reversible C-automaton \\
$A=(Q,\Sigma,q_0,Q_F,\delta)$ is specified by $Q$, $\Sigma$,
$q_0$, $Q_F$ defined as in 1-way PRA-C Definition \ref{prac-def},
and a transition function $$ \delta:Q\times\Gamma\times Q\times D
\longrightarrow\bbbr_{[0,1]}, $$ where $\Gamma$ defined as in
1-way PRA-C definition and $D=\{0,1\}$ denotes whether automaton
stays on the same position or moves one letter ahead on the input
tape. Furthermore, transition function satisfies the following
requirements:

\begin{eqnarray}
&&\forall(q_1,\sigma_1)\in Q\times\Gamma\ \sum\limits_{q\in Q,d\in
D}\delta(q_1,\sigma_1,q,d)=1 \label{sc3-2-2}\\
&&\forall(q_1,\sigma_1,\sigma_2)\in Q\times\Gamma^2\ \sum\limits_{q\in Q
}\delta(q,\sigma_1,q_1,0)+\sum\limits_{q\in Q, \sigma \in \Gamma
}\delta(q,\sigma_2,q_1,1)=1 \label{sc3-1-2}
\end{eqnarray}

\end{definition}

\begin{theorem}
Language (a,b)*a is recognizable by 1.5-way weakly reversible PRA-C.
\end{theorem}

\begin{proof}

The $Q=\{q_{0},q_{1}\}$, $Q_{F}=\{q_{1}\}$, $\delta$ is defined as
follows

$\begin{array}{cccc}
  \delta(q_{0},a,q_{0},0)=\frac{1}{2} & \delta(q_{0},a,q_{1},1)=\frac{1}{2} &
  \delta(q_{1},a,q_{0},0)=\frac{1}{2} & \delta(q_{1},a,q_{1},1)=\frac{1}{2} \\
  \delta(q_{0},b,q_{0},1)=\frac{1}{2} & \delta(q_{0},b,q_{1},0)=\frac{1}{2} &
  \delta(q_{1},b,q_{0},1)=\frac{1}{2} & \delta(q_{1},b,q_{1},0)=\frac{1}{2} \\
   \delta(q_{0},\$,q_{0},1)=1  &  \delta(q_{1},\$,q_{1},1)=1
\end{array}$

It is easy to check that such automaton moves ahead according to the
transition of the following deterministic automaton

$\begin{array}{cc}
  \delta(q_{0},a,q_{1},1)=1 & \delta(q_{1},a,q_{1},1)=1 \\
  \delta(q_{0},b,q_{0},1)=1 & \delta(q_{1},b,q_{0},1)=1 \\
   \delta(q_{0},\$,q_{0},1)=1  &  \delta(q_{1},\$,q_{1},1)=1
\end{array}$
%\end{proof}

So the probability of wrong answer is 0. The probability to be at
the $m$-th position of the input tape after $n$ steps of calculation
for $m \leq n$ is $C^{m}_{n} $. Therefore it is necessary no more
than $O(n*\log(p))$ steps to reach the end of the word of length n
(and so obtain correct answer) with probability $1-\frac{1}{p}$.
\qed
\end{proof}

\section{A Classification of Reversible Automata}
\label{hierarhija}
We propose the following classification for finite 1-way reversible automata:\\
\hbadness=10000
\vbadness=10000
\begin{tabular}{|p{2.5cm}|p{4.5cm}|p{4.5cm}|}
\hline
& C-automata & DH-automata\\
\hline
Deterministic Automata &
\mbox{Permutation Automata} \cite{HS 66,T 68} (DRA-C) &
Reversible Finite Automata \cite{AF 98} (DRA-DH)\\
\hline
Quantum \mbox{Automata} with Pure States &
\mbox{Measure-Once Quantum} \mbox{Finite Automata \cite{MC 00}} (QRA-P-C)&
\mbox{Measure-Many Quantum} \mbox{Finite Automata \cite{KW 97}} (QRA-P-DH)\\
\hline
Probabilistic Automata &
\mbox{Probabilistic Reversible} \mbox{C-automata} (PRA-C)&
\mbox{Probabilistic Reversible} \mbox{DH-automata} (PRA-DH)\\
\hline
Quantum Finite Automata with
Mixed States &
\mbox{not considered yet} \mbox{(QRA-M-C)}&
\mbox{Enhanced Quantum} \mbox{Finite Automata \cite{N 99}} \mbox{(QRA-M-DH)}\\
\hline
\end{tabular}
\hbadness=1000
\vbadness=1000

Language class problems are solved for DRA-C, DRA-DH,
QRA-P-C, for the rest types they are still open.
Every type of DH-automata may simulate the corresponding type of C-automata.

Generally, language classes recognized by C-automata are closed under\\
boolean operations
(though this is open for QRA-M-C), while DH-automata are not (though this is open for QRA-M-DH and possibly for
PRA-DH).

\begin{definition}
\label{proto}
We say that a unitary matrix $U$ is a {\em prototype} for a doubly stochastic matrix $S$,
if $\forall i,j$ $|U_{i,j}|^2=S_{i,j}$.
\end{definition}

Not every doubly stochastic matrix has a unitary prototype.
Such matrix is, for example,
$\left(
\begin{array}{ccc}
\frac{1}{2} & \frac{1}{2} & 0\\
\frac{1}{2} & 0 & \frac{1}{2}\\
0 & \frac{1}{2} & \frac{1}{2}\\
\end{array}
\right)$.

In Introduction, we demonstrated some relation between PRA-C and QRA-M-DH (and hence, QRA-M-C).
However, due to the example above, we do not know exactly, whether every PRA-C can be simulated by QRA-M-C, or whether
every PRA-DH can be simulated by QRA-M-DH.

\begin{theorem}
If all matrices of a PRA-C have unitary prototypes, then the PRA-C may be simulated by a QRA-M-C and by a QRA-M-DH.
\end{theorem}
\begin{proof}
Trivial.
\qed
\end{proof}

\begin{theorem}
If all matrices of a PRA-DH have unitary prototypes, then the PRA-DH may be simulated by a QRA-M-DH.
\end{theorem}
\begin{proof}
Trivial.
\qed
\end{proof}

\appendix

\section{End-Marker Theorems for PRA-C Automata}
We denote a PRA-C with both end-markers as \#,\$-PRA-C.
We denote a PRA-C with left end-marker only as \#-PRA-C.

\begin{theorem}
Let $A$ be a \#,\$-PRA-C, which recognizes a language $L$. There exists a
\#-PRA-C which recognizes the same language.
\end{theorem}
\begin{proof}
Suppose $A=(Q,\Sigma,q_0,Q_F,\delta)$, where $|Q|=n$. $A$ recognizes $L$ with interval $(p_1,p_2)$.
We construct the following automaton
$A^\prime=(Q^\prime,\Sigma,q_{0,0},Q^\prime_F,\delta^\prime)$ with $mn$ states.
Informally, $A^\prime$ equiprobably simulates $m$ copies of the automaton $A$.

$Q^\prime=\{q_{0,0},\ldots,q_{0,m-1},q_{1,0},\ldots,q_{1,m-1},\ldots,q_{n-1,0},\ldots,q_{n-1,m-1}\}$.

If $\sigma\neq \#$, $\delta^\prime(q_{i,k},\sigma,q_{j,l})=\left\{
\begin{array}{l}
\delta(q_i,\sigma,q_j)\mbox{, if }k=l\\
0\mbox{, if }k\neq l.\\
\end{array}\right.$
\\Otherwise, $\delta^\prime(q_{0,0},\#,q_{j,l})=\frac{1}{m}\delta(q_0,\#,q_j)$, and if $q_{i,k}\neq q_{0,0}$,
$\delta^\prime(q_{i,k},\#,q)=\frac{1-\delta^\prime(q_{0,0},\#,q)}{mn-1}$.
Function $\delta^\prime$ satisfies the requirements (\ref{sc1-1}) and (\ref{sc1-2}) of Definition \ref{prac-def}.

We define $Q^\prime_F$ as follows. A state $q_{i,k}\in Q^\prime_F$ if and only if $0\leq k<mp(q_i)$, where 
$p(q_i)\stackrel{\scriptscriptstyle {\rm def}}{=}\sum\limits_{q\in Q_F}\delta(q_i,\$,q)$.

Suppose $\#\omega\$$ is an input word. Having read $\#\omega$, $A$ is in superposition
$\sum\limits_{i=0}^{n-1}a^\omega_iq_i$. After $A$ has read $\$$, $\#\omega\$$ is accepted with
probability $p_\omega=\sum\limits_{i=0}^{n-1}a^\omega_ip(q_i)$.

On the other hand, having read $\#\omega$, $A^\prime$ is in superposition
$\frac{1}{m}\sum\limits_{j=0}^{m-1}\sum\limits_{i=0}^{n-1}a^\omega_iq_{i,j}$. So the input word $\#\omega$
is accepted with probability $p^\prime_\omega=\frac{1}{m}\sum\limits_{i=0}^{n-1}a^\omega_i\lceil mp(q_i)\rceil$.

Consider $\omega\in L$. Then 
$p^\prime_\omega=\frac{1}{m}\sum\limits_{i=0}^{n-1}a^\omega_i\lceil mp(q_i)\rceil
\geq\sum\limits_{i=0}^{n-1}a_i^\omega p(q_i)=p_\omega\geq p_2$.

Consider $\xi\notin L$. Then 
$p^\prime_\xi=\frac{1}{m}\sum\limits_{i=0}^{n-1}a^\xi_i\lceil mp(q_i)\rceil
<\sum\limits_{i=0}^{n-1}a_i^\xi p(q_i)+\frac{1}{m}\sum\limits_{i=0}^{n-1}a_i^\xi=
p_\xi+\frac{1}{m}\leq p_1+\frac{1}{m}$.

Therefore $A^\prime$ recognizes $L$ with bounded error, provided $m>\frac{1}{p_2-p_1}$.
\qed
\end{proof}

Now we are going to prove that PRA-C without end-markers recognize the same languages as \#-PRA-C automata.

If $A$ is a \#-PRA-C, then, having read the left end-marker $\#$, the automaton simulates some other automata
$A_0,A_1,\ldots,A_{m-1}$ with positive probabilities $p_0,\ldots,p_{m-1}$, respectively.
$A_0,A_1,\ldots,A_{m-1}$ are automata without end-markers. By $p_{i,\omega}$, $0\leq i<m$, we denote the probability
that the automaton $A_i$ accepts the word $\omega$.

We prove the following lemma first.

\begin{lemma}
\label{A.2}
Suppose $A^\prime$ is a \#-PRA-C which recognizes a language $L$ with interval $(a_1,a_2)$.
Then for every $\varepsilon$, $0<\varepsilon<1$, exists a \#-PRA-C $A$ which recognizes $L$ with interval $(a_1,a_2)$,
such that
\begin{itemize}
\item[a)] if $\omega\in L$, $p_{0,\omega}+p_{1,\omega}+\ldots+p_{n-1,\omega}>\frac{a_2n}{1+\varepsilon}$
\item[b)] if $\omega\notin L$, $p_{0,\omega}+p_{1,\omega}+\ldots+p_{n-1,\omega}<\frac{a_1n}{1-\varepsilon}$.
\end{itemize}
Here $n$ is the number of automata without end-markers, being simulated by $A$, and $p_{i,\omega}$ is
the probability that i-th simulated automaton $A_i$ accepts $\omega$.
\end{lemma}
\begin{proof}
Suppose a \#-PRA-C $A^\prime$ recognizes a language $L$ with interval $(a_1,a_2)$. 
Having read the symbol $\#$, $A^\prime$ simulates automata $A_0^\prime,\ldots,A_{m-1}^\prime$ with
probabilities $p_0^\prime,\ldots,p_{m-1}^\prime$, respectively. We choose $\varepsilon$, $0<\varepsilon<1$.

By Dirichlet's principle (\cite{HW 79}, p. 170), $\forall\varphi>0$ exists $n\in\bbbn^+$
such that $\forall i$ $p_i^\prime n$ differs from some positive integer by less than $\varphi$.

Let $0<\varphi<\min\left(\frac{1}{m},\varepsilon\right)$. Let $g_i$ be the nearest integer of $p_i^\prime n$.
So $|p_i^\prime n-g_i|<\varphi$ and 
$\left|\frac{p_i^\prime}{g_i}-\frac{1}{n}\right|<\frac{\varphi}{ng_i}\leq\frac{\varphi}{n}$. Since
$|p_i^\prime n-g_i|<\varphi$,
we have 
$\left|n-\sum\limits_{i=0}^{m-1}g_i\right|<\varphi m<1$. Therefore,
since $g_i\in\bbbn^+$, $\sum\limits_{i=0}^{m-1}g_i=n$.

Now we construct the \#-PRA-C $A$, which satisfies the properties expressed in Lemma \ref{A.2}.
For every $i$, we make $g_i$ copies of $A_i^\prime$. Having read $\#$, for every $i$
$A$ simulates each copy of $A_i^\prime$ with
probability $\frac{p_i^\prime}{g_i}$. The construction of $V_\#$ is equivalent to that used in the
proof of Lemma \ref{union-lemma}. Therefore $A$ is characterized by doubly stochastic matrices.
$A$ recognizes $L$ with the same interval as $A^\prime$, i.e., $(a_1,a_2)$.

Using new notations, $A$ simulates $n$ automata
$A_0,A_1,\ldots,A_{n-1}$ with
probabilities $p_0,p_1,\ldots,p_{n-1}$, respectively. Note that 
$\forall i\ \left|p_i-\frac{1}{n}\right|<\frac{\varphi}{n}$.
Let $p_{i,\omega}$ be the probability that $A_i$ accepts the word $\omega$.

Consider $\omega\in L$. We have $p_0p_{0,\omega}+p_1p_{1,\omega}+\ldots+p_{n-1}p_{n-1,\omega}\geq a_2$.
Since $p_i<\frac{1+\varphi}{n}$,
$\frac{1+\varphi}{n}(p_{0,\omega}+p_{1,\omega}+\ldots+p_{n-1,\omega})>a_2$.
Hence $$p_{0,\omega}+p_{1,\omega}+\ldots+p_{n-1,\omega}>\frac{a_2n}{1+\varphi}>\frac{a_2n}{1+\varepsilon}.$$

Consider $\xi\notin L$. We have $p_0p_{0,\xi}+p_1p_{1,\xi}+\ldots+p_{n-1}p_{n-1,\xi}\leq a_1$.
Since $p_i>\frac{1-\varphi}{n}$,
$\frac{1-\varphi}{n}(p_{0,\xi}+p_{1,\xi}+\ldots+p_{n-1,\xi})<a_1$.
% Therefore
%$\frac{1}{n}(p_{0,\xi}+p_{1,\xi}+\ldots+p_{n-1,\xi})<\frac{\varepsilon_1}{1-\varphi}<
%\frac{\frac{\varepsilon}{2}}{1-\frac{\varepsilon}{2}}<\varepsilon$. 
Hence
$$p_{0,\xi}+p_{1,\xi}+\ldots+p_{n-1,\xi}<\frac{a_1n}{1-\varphi}<\frac{a_1n}{1-\varepsilon}.$$
\qed
\end{proof}

\begin{theorem}
Let $A$ be a \#-PRA-C, which recognizes a language $L$. There exists a PRA-C without end-markers,
which recognizes the same language.
\end{theorem}
\begin{proof}
Consider a \#-PRA-C which recognizes a language $L$ with interval $(a_1,a_2)$. Using Lemma \ref{A.2},
we choose $\varepsilon$, $0<\varepsilon<\frac{a_2-a_1}{a_2+a_1}$,
and construct an automaton $A^\prime$ which recognizes L with interval $(a_1,a_2)$, with the following
properties.

Having read $\#$, $A^\prime$ simulates $A_0^\prime,\ldots,A_{m-1}^\prime$
with probabilities $p_0^\prime,\ldots,p_{m-1}^\prime$, respectively.
$A_0^\prime,\ldots,A_{m-1}^\prime$ are
automata without end-markers. $A_i^\prime$ accepts $\omega$ with probability $p_{i,\omega}^\prime$.
If $\omega\in L$,
$p_{0,\omega}^\prime+p_{1,\omega}^\prime+\ldots+p_{m-1,\omega}^\prime>\frac{a_2m}{1+\varepsilon}$.
Otherwise, if 
$\omega\notin L$, $p_{0,\omega}^\prime+p_{1,\omega}^\prime+\ldots+p_{m-1,\omega}^\prime<\frac{a_1m}{1-\varepsilon}$.

That also implies that for every $n=km$, $k\in\bbbn^+$, we are able to construct a \#-PRA-C $A$ which recognizes
$L$ with interval $(a_1,a_2)$, such that
\begin{itemize}
\item[a)] if 
$\omega\in L$, $p_{0,\omega}+p_{1,\omega}+\ldots+p_{n-1,\omega}>\frac{a_2n}{1+\varepsilon}$;
\item[b)] if 
$\omega\notin L$, $p_{0,\omega}+p_{1,\omega}+\ldots+p_{n-1,\omega}<\frac{a_1n}{1-\varepsilon}$.
\end{itemize}

$A$ simulates $A_0,\ldots,A_{n-1}$. Let us consider the system $F_n=(A_0,\ldots,A_{n-1})$.
Let $\delta=\frac{1}{2}(a_1+a_2)$. Since
$\varepsilon<\frac{a_2-a_1}{a_2+a_1}$, $\frac{a_2}{1+\varepsilon}>\delta$ and
$\frac{a_1}{1-\varepsilon}<\delta$. As in the proof of Theorem
\ref{1eps-lem}, we define that the system accepts a word, if more than $n\delta$ automata in the system
accept the word.

Let us take $\eta_0$, such that
%$0<\eta_0<\min\left(\frac{a_2}{1+\varepsilon}-\delta,\delta-\frac{a_1}{1-\varepsilon}\right)$.
$0<\eta_0<\frac{a_2}{1+\varepsilon}-\delta<\delta-\frac{a_1}{1-\varepsilon}$.

Consider $\omega\in L$. We have that 
$\sum\limits_{i=0}^{n-1}p_{i,\omega}>\frac{a_2n}{1+\varepsilon}>n\delta$. As a result
of reading $\omega$, $\mu_n^\omega$ automata in the system accept the word, and the rest reject it.
The system has accepted the word, if $\frac{\mu_n^\omega}{n}>\delta$.
Since
$0<\eta_0<\frac{a_2}{1+\varepsilon}-\delta<\frac{1}{n}\sum\limits_{i=0}^{n-1}p_{i,\omega}-\delta$,
we have 
\begin{equation}
\label{dd}
P\left\{\frac{\mu_n^\omega}{n}>\delta\right\}\geq 
P\left\{\left|\frac{\mu_n^\omega}{n}-\frac{1}{n}\sum\limits_{i=0}^{n-1}p_{i,\omega}\right|<\eta_0\right\}.
\end{equation}
If we look on $\frac{\mu_n^\omega}{n}$ as a random variable $X$, $E(X)=\frac{1}{n}\sum\limits_{i=0}^{n-1}p_{i,\omega}$
and variance $V(X)=\frac{1}{n^2}\sum\limits_{i=0}^{n-1}p_{i,\omega}(1-p_{i,\omega})$, therefore
Chebyshev's inequality yields the following:
$$
P\left\{\left|\frac{\mu_n^\omega}{n}-\frac{1}{n}\sum\limits_{i=0}^{n-1}p_{i,\omega}\right|\geq\eta_0\right\}\leq
\frac{1}{n^2\eta_0^2}\sum\limits_{i=0}^{n-1}p_{i,\omega}(1-p_{i,\omega})\leq\frac{1}{4n\eta_0^2}.
$$
That is equivalent to
$P\left\{\left|\frac{\mu_n^\omega}{n}-\frac{1}{n}\sum\limits_{i=0}^{n-1}p_{i,\omega}\right|<\eta_0\right\}
\geq1-\frac{1}{4n\eta_0^2}$.
So, taking into account (\ref{dd}), 
\begin{equation}
\label{kk}
P\left\{\frac{\mu_n^\omega}{n}>\delta\right\}\geq1-\frac{1}{4n\eta_0^2}.
\end{equation}

On the other hand, consider $\xi\notin L$. So $\sum\limits_{i=0}^{n-1}p_{i,\xi}<\frac{a_1n}{1-\varepsilon}<n\delta$.
Again, since
$0<\eta_0<\delta-\frac{a_1}{1-\varepsilon}<\delta-\frac{1}{n}\sum\limits_{i=0}^{n-1}p_{i,\xi}$,
\begin{equation}
\label{ee}
P\left\{\frac{\mu_n^\xi}{n}>\delta\right\}\leq 
P\left\{\left|\frac{\mu_n^\xi}{n}-\frac{1}{n}\sum\limits_{i=0}^{n-1}p_{i,\xi}\right|
\geq\eta_0\right\}\leq\frac{1}{4n\eta_0^2}.
\end{equation}

The constant $\eta_0$ does not depend on $n$ and $n$ may be chosen sufficiently large. Therefore,
by (\ref{kk}) and (\ref{ee}), the system $F_n$ recognizes $L$ with bounded error, if $n>\frac{1}{2\eta_0^2}$.

Following a way identical to that used in the proof of Theorem \ref{1eps-lem},
it is possible to construct a single PRA-C without end-markers, which simulates
the system $F_n$ and therefore recognizes the language $L$.
\qed
\end{proof}

\end{document}